\documentclass[auto]{article}
\usepackage{supertabular,lscape,epsfig}
\usepackage{amssymb}
\usepackage{amsmath}
\usepackage{xcolor}
\usepackage{rotating}
\DeclareSymbolFont{ppa}{OT1}{ppl}{m}{it}
\DeclareMathSymbol{\vv}{\mathalpha}{ppa}{'166}

\begin{document}

\newcommand{\dd}{\,{\rm d}}
\newcommand{\ie}{{\it i.e.},\,}
\newcommand{\etal}{{\it et al.\ }}
\newcommand{\eg}{{\it e.g.},\,}
\newcommand{\cf}{{\it cf.\ }}
\newcommand{\vs}{{\it vs.\ }}
\newcommand{\zdot}{\makebox[0pt][l]{.}}
\newcommand{\up}[1]{\ifmmode^{\rm #1}\else$^{\rm #1}$\fi}
\newcommand{\dn}[1]{\ifmmode_{\rm #1}\else$_{\rm #1}$\fi}
\newcommand{\upd}{\up{d}}
\newcommand{\uph}{\up{h}}
\newcommand{\upm}{\up{m}}
\newcommand{\ups}{\up{s}}
\newcommand{\arcd}{\ifmmode^{\circ}\else$^{\circ}$\fi}
\newcommand{\arcm}{\ifmmode{'}\else$'$\fi}
\newcommand{\arcs}{\ifmmode{''}\else$''$\fi}
\newcommand{\MS}{{\rm M}\ifmmode_{\odot}\else$_{\odot}$\fi}
\newcommand{\RS}{{\rm R}\ifmmode_{\odot}\else$_{\odot}$\fi}
\newcommand{\LS}{{\rm L}\ifmmode_{\odot}\else$_{\odot}$\fi}
\newcommand{\feh}{\hbox{$ [{\rm Fe}/{\rm H}]$}}

\newcommand{\Abstract}[2]{{\footnotesize\begin{center}ABSTRACT\end{center}
\vspace{1mm}\par#1\par
\noindent
{~}{\it #2}}}

\newcommand{\TabCap}[2]{\begin{center}\parbox[t]{#1}{\begin{center}
  \small {\spaceskip 2pt plus 1pt minus 1pt T a b l e}
  \refstepcounter{table}\thetable \\[2mm]
  \footnotesize #2 \end{center}}\end{center}}

\newcommand{\TableSep}[2]{\begin{table}[p]\vspace{#1}
\TabCap{#2}\end{table}}

\newcommand{\FigCap}[1]{\footnotesize\par\noindent Fig.\  %
  \refstepcounter{figure}\thefigure. #1\par}

\newcommand{\TableFont}{\footnotesize}
\newcommand{\TableFontIt}{\ttit}
\newcommand{\SetTableFont}[1]{\renewcommand{\TableFont}{#1}}

\newcommand{\MakeTable}[4]{\begin{table}[htb]\TabCap{#2}{#3}
  \begin{center} \TableFont \begin{tabular}{#1} #4
  \end{tabular}\end{center}\end{table}}

\newcommand{\MakeTableSep}[4]{\begin{table}[p]\TabCap{#2}{#3}
  \begin{center} \TableFont \begin{tabular}{#1} #4
  \end{tabular}\end{center}\end{table}}

\newenvironment{references}%
{
\footnotesize \frenchspacing
\renewcommand{\thesection}{}
\renewcommand{\in}{{\rm in }}
\renewcommand{\AA}{Astron.\ Astrophys.}
\newcommand{\AAS}{Astron.~Astrophys.~Suppl.~Ser.}
\newcommand{\ApJ}{Astrophys.\ J.}
\newcommand{\ApJS}{Astrophys.\ J.~Suppl.~Ser.}
\newcommand{\ApJL}{Astrophys.\ J.~Letters}
\newcommand{\AJ}{Astron.\ J.}
\newcommand{\IBVS}{IBVS}
\newcommand{\PASJ}{PASJ}
\newcommand{\PASP}{P.A.S.P.}
\newcommand{\Acta}{Acta Astron.}
\newcommand{\MNRAS}{MNRAS}

\renewcommand{\and}{{\rm and }}
\section{{\rm REFERENCES}}
\sloppy \hyphenpenalty10000
\begin{list}{}{\leftmargin1cm\listparindent-1cm
\itemindent\listparindent\parsep0pt\itemsep0pt}}%
{\end{list}\vspace{2mm}}

\def\TYLDA{~}
\newlength{\DW}
\settowidth{\DW}{0}
\newcommand{\dw}{\hspace{\DW}}

\newcommand{\refitem}[5]{\item[]{#1} #2%
\def\REFARG{#3}\ifx\REFARG\TYLDA\else, {\it#3}\fi
\def\REFARG{#4}\ifx\REFARG\TYLDA\else, {\bf#4}\fi
\def\REFARG{#5}\ifx\REFARG\TYLDA\else, {#5}\fi.}

\newcommand{\Section}[1]{\section{#1}}
\newcommand{\Subsection}[1]{\subsection{#1}}
\newcommand{\Acknow}[1]{\par\vspace{5mm}{\bf Acknowledgments.} #1}
\pagestyle{myheadings}

\newfont{\bb}{ptmbi8t at 12pt}
\newcommand{\xrule}{\rule{0pt}{2.5ex}}
\newcommand{\xxrule}{\rule[-1.8ex]{0pt}{4.5ex}}
\def\thefootnote{\fnsymbol{footnote}}
\newcommand{\ml}[1]{\textcolor{red}{#1}}

\begin{center}
{\Large\bf Spectroscopic Observations of Four Candidates for Blue Large-Amplitude Pulsators. No BLAPs at High Galactic Latitudes\footnote{Based on observations obtained with the 6.5-m Magellan/Baade telescope at Las Campanas Observatory of the Carnegie Institution for Science}}
\vskip1cm
{\bf
P.~~P~i~e~t~r~u~k~o~w~i~c~z$^1$,~~M.~~L~a~t~o~u~r$^2$,~~F.~~D~i~~M~i~l~l~e$^3$,~~M.~~D~o~r~s~c~h$^4$,\\~~U.~~H~e~b~e~r$^5$,~~and~~J.~~B~o~r~o~w~i~c~z$^1$\\}
\vskip3mm
{
$^1$ Astronomical Observatory, University of Warsaw, Al. Ujazdowskie 4, 00-478 Warszawa, Poland\\
$^2$ Institute for Astrophysics, Georg-August-University G\"ottingen, Friedrich-Hund-Platz 1, D-37077 G\"ottingen, Germany\\
$^3$ Las Campanas Observatory, Carnegie Institution for Science, Colina el Pino, Casilla 601, La Serena, Chile\\
$^4$ Institut f\"ur Physik und Astronomie, Universit\"at Potsdam, Haus 28, Karl-Liebknecht-Str. 24/25, 14476 Potsdam, Germany\\
$^5$ Dr. Remeis-Sternwarte \& ECAP, Friedrich-Alexander University, Erlangen-N\"urnberg, Sternwartstr. 7, 96049 Bamberg, Germany\\
}
\end{center}

\Abstract{
We have obtained spectroscopic observations for four short-period variable objects detected in ZTF, Gaia, and Pan-STARRS (ZGP) data and classified as Blue Large-Amplitude Pulsators (BLAPs) in McWhirter and Lam (2022): ZGP-BLAP-03, ZGP-BLAP-04, ZGP-BLAP-10, and ZGP-BLAP-15. The variables have periods between 46 and 56~min, full amplitudes of 0.13-0.22~mag in the $r$ band, and light curve shapes typical for radially pulsating stars. Three of them were found at high galactic latitudes ($|b|>30\arcd$). We have identified object ZGP-BLAP-03 as an early F-type star, while objects ZGP-BLAP-04 and ZGP-BLAP-15 as low-metallicity late A-type stars. These are the three objects found at high galactic latitudes and located several kiloparsecs from the Sun. Thus, they are SX Phoenicis-type variable stars residing in the Galactic halo. In the case of low-latitude object ZGP-BLAP-10, we report the presence of helium lines in its spectrum and atmospheric parameters in agreement with known BLAPs. This and other results indicate that BLAPs are absent in metal-poor environments.}

{Stars: oscillations – Stars: variables: general}


\Section{Introduction}

Blue Large-Amplitude Pulsators (BLAPs) are a class of hot pulsating variable stars exhibiting exceptionally large light variations with amplitudes of 0.1--0.4 mag at a short period of 3--75 min. Most of the variables have characteristic sawtooth-shaped light curves. In some longer-period BLAPs, there is an additional bump. BLAPs have effective temperatures ($T_{\rm eff}$) around 30~000~K, surface gravities ($\log g$) of 4.2--5.7 dex (in cgs units), and a helium-to-hydrogen content $\log (N_{\rm He}/N_{\rm H})$ either around $-2.1$ dex or around $-0.6$ dex (Pietrukowicz \etal 2025).

Theoretical models show that enhanced metallicity (high iron and nickel abundance) in the envelope of the star is necessary to drive BLAP pulsations (Byrne and Jeffery 2020, Jeffery 2025). The following three models for the inner structure of BLAPs have been suggested:
\begin{itemize}
\item BLAPs are low-mass ($M \sim 0.3 \MS$) helium-core, shell-hydrogen-burning pre-white dwarfs (Romero \etal 2018, Byrne and Jeffery 2018).
\item BLAPs are $M \sim 0.45 \MS$ helium-core burning post-giants evolving toward the extended horizontal branch (EHB) (Wu and Li 2018).
\item BLAPs are $M \gtrsim 0.45 \MS$ stars with helium-burning shell above the CO core evolving off the EHB (Xiong \etal 2022, Lin \etal 2023).
\end{itemize}
The following hypotheses regarding the origin of BLAPs have been proposed:
\begin{itemize}
\item BLAPs are giants that were stripped off during a close encounter with the supermassive black hole in the center of the Milky Way (Pietrukowicz \etal 2017).
\item BLAPs are surviving companions from single-degenerate Type Ia supernova explosions (Meng \etal 2020).
\item BLAPs could form via mass transfer or common envelope ejection in binary system evolution (Byrne \etal 2021).
\item BLAPs are the result of a merger between a helium-core white dwarf and a low-mass main-sequence star (Zhang \etal 2023).
\item BLAPs are the result of a merger of CO white dwarfs (Pigulski \etal 2024).
\end{itemize}

The first BLAP was discovered in the Galactic disk at coordinates $(l,b)=(288\zdot\arcd 06, -2\zdot\arcd 35)$ (Pietrukowicz \etal 2013) in the data from the third phase of the Optical Gravitational Lensing Experiment (OGLE; Udalski \etal 2008). The detection of thirteen very similar variables in the Galactic bulge fields of the fourth phase of the OGLE survey (Udalski \etal 2015) allowed the announcement of the new class of hot pulsating stars, the Blue Large-Amplitude Pulsators (Pietrukowicz \etal 2017). Subsequently, four BLAPs with extremely short periods of several minutes and an order of magnitude stronger surface gravities were discovered in the Zwicky Transient Facility (ZTF) data (Kupfer \etal 2019). An additional four genuine pulsators were detected in the OmegaWhite (Ramsay \etal 2022) and SkyMapper data (Chang \etal 2024). Very recently, nearly 140 new BLAPs have been discovered in the entire OGLE survey area covering almost 3000 square degrees of the Milky Way stripe visible from Las Campanas Observatory, Chile (Borowicz \etal 2023ab, Pietrukowicz \etal 2025, Borowicz \etal in prep.), increasing the number of known such pulsators to around 200. Twenty-two BLAP candidates were reported in McWhirter and Lam (2022), four of which are the subject of the present work. In spite of the growing list of BLAPs and BLAP candidates identified from their photometric variability, only 28 of them were confirmed spectroscopically before this work. They are listed in Table~1.

\begin{sidewaystable}
\caption{Spectroscopically confirmed BLAPs}
\label{}
\begin{center}
{\scriptsize
\begin{tabular}{lllccccccccc}
\hline
Object name                & RA(2000.0)  &  Dec(2000.0)  &  $G$  &  $P$   & Year & Spectrograph/  & Sp.   & $T_{\rm eff}$  & $\log g$      & $\log(N_{\rm He}/N_{\rm H})$ & Ref. \\
                           &             &               & [mag] & [min]  & of obs. &  /Telescope & res.  &      [K]       & [cm s$^{-2}$] & & \\
\hline
OGLE-BLAP-001              & 10:41:48.77 & $-61$:25:08.5 & 17.48 & 28.255 & 2016 & MagE/Magellan  & 4100  & $30~800\pm 500$ & $4.61\pm0.07$ & $-0.55\pm0.05$ & [1] \\
OGLE-BLAP-009              & 17:58:48.20 & $-27$:16:53.7 & 15.55 & 31.935 & 2016 & GMOS/Gemini-S  &  800  & $31~800\pm1400$ & $4.40\pm0.18$ & $-0.41\pm0.13$ & [1,2] \\
OGLE-BLAP-011              & 18:00:23.24 & $-35$:58:03.1 & 17.22 & 34.875 & 2016 & GMOS/Gemini-S  &  800  & $26~200\pm2900$ & $4.20\pm0.29$ & $-0.45\pm0.11$ & [1] \\
OGLE-BLAP-014              & 18:12:41.79 & $-31$:12:07.8 & 16.75 & 33.623 & 2016 & GMOS/Gemini-S  &  800  & $30~900\pm2100$ & $4.42\pm0.26$ & $-0.54\pm0.16$ & [1] \\
OW-BLAP-1                  & 18:12:27.88 & $-29$:38:48.4 & 18.26 & 10.854 & 2016 & RSS/SALT       & 2000  & $30~600\pm2500$ & $4.67\pm0.25$ & $-2.1 \pm0.2 $ & [3] \\
OW-BLAP-2                  & 18:11:00.22 & $-27$:30:13.3 & 17.89 & 22.803 & 2016 & RSS/SALT       & 2000  & $27~300\pm1500$ & $4.83\pm0.20$ & $-0.7 \pm0.1 $ & [3] \\
OW-BLAP-$3^*$              & 18:10:38.58 & $-25$:16:08.7 & 18.95 & 28.975 & 2016 & RSS/SALT       &  900  & $29~900\pm3500$ & $4.16\pm0.40$ & $-0.8 \pm0.3 $ & [3] \\
ZTF-HG-BLAP-1              & 07:13:29.02 & $-15$:21:25.2 & 16.51 &  3.337 & 2018 & LRIS/Keck-I    & 1000  & $34~000\pm 500$ & $5.70\pm0.05$ & $-2.1 \pm0.2 $ & [4] \\
ZTF-HG-BLAP-2              & 18:45:21.40 & $-25$:44:37.5 & 18.81 &  6.053 & 2018 & LRIS/Keck-I    & 1000  & $31~400\pm 600$ & $5.41\pm0.06$ & $-2.2 \pm0.3 $ & [4] \\
ZTF-HG-BLAP-3              & 19:13:06.79 & $-12$:05:44.6 & 17.58 &  7.314 & 2018 & LRIS/Keck-I    & 1000  & $31~600\pm 600$ & $5.33\pm0.05$ & $-2.0 \pm0.2 $ & [4] \\
ZTF-HG-BLAP-4              & 18:28:15.88 & $+12$:25:30.5 & 17.25 &  7.925 & 2018 & DBSP/200" Hale & 1500  & $31~700\pm 500$ & $5.31\pm0.05$ & $-2.4 \pm0.4 $ & [4] \\
OGLE-BLAP-010              & 17:58:59.22 & $-35$:18:07.0 & 17.15 & 32.133 & 2018 & MagE/Magellan  & 4100  & $29~800\pm 300$ & $4.57\pm0.04$ & $-0.58\pm0.03$ & [5] \\
OGLE-BLAP-021              & 17:58:30.23 & $-29$:38:09.2 & 17.80 & 42.742 & 2018 & MagE/Magellan  & 4100  & $28~500\pm 300$ & $4.46\pm0.04$ & $-0.64\pm0.03$ & [5] \\
OGLE-BLAP-022              & 17:58:51.29 & $-28$:28:53.3 & 17.86 & 74.052 & 2018 & MagE/Magellan  & 4100  & $28~900\pm 400$ & $4.45\pm0.06$ & $-0.74\pm0.05$ & [5] \\
OGLE-BLAP-030              & 18:04:57.10 & $-28$:08:06.2 & 17.97 & 21.160 & 2018 & MagE/Magellan  & 4100  & $31~400\pm 300$ & $4.85\pm0.05$ & $-0.75\pm0.04$ & [5] \\
OGLE-BLAP-031              & 18:05:56.03 & $-35$:53:55.5 & 17.16 & 40.671 & 2018 & MagE/Magellan  & 4100  & $26~800\pm 200$ & $4.38\pm0.03$ & $-0.54\pm0.03$ & [5] \\
OGLE-BLAP-033              & 18:07:31.70 & $-27$:58:20.8 & 18.87 & 15.822 & 2018 & MagE/Magellan  & 4100  & $33~100\pm 700$ & $5.04\pm0.11$ & $-0.88\pm0.07$ & [5] \\
OGLE-BLAP-034              & 18:07:50.55 & $-30$:04:47.6 & 17.55 & 44.335 & 2018 & MagE/Magellan  & 4100  & $30~300\pm 300$ & $4.49\pm0.04$ & $-0.62\pm0.03$ & [5] \\
OGLE-BLAP-037              & 18:19:20.89 & $-27$:29:56.3 & 17.19 & 15.712 & 2018 & MagE/Magellan  & 4100  & $32~800\pm 200$ & $4.93\pm0.04$ & $-2.15\pm0.05$ & [5] \\
OGLE-BLAP-019              & 17:53:34.39 & $-30$:02:11.6 & 18.05 & 48.005 & 2019 & MagE/Magellan  & 4100  & $28~000\pm 700$ & $4.29\pm0.09$ & $-0.76\pm0.08$ & [5] \\
OGLE-BLAP-020              & 17:56:32.05 & $-20$:37:15.7 & 18.29 & 47.923 & 2019 & MagE/Magellan  & 4100  & $29~200\pm 500$ & $4.40\pm0.07$ & $-0.59\pm0.06$ & [5] \\
OGLE-BLAP-024              & 17:59:43.70 & $-28$:44:26.4 & 17.28 & 48.268 & 2019 & MagE/Magellan  & 4100  & $25~200\pm 300$ & $4.39\pm0.05$ & $-0.66\pm0.04$ & [5] \\
SMSS J184506.82-300804.7$^\dag$ & 18:45:06.82 & $-30$:08:04.7 & 16.48 & 20.168 & 2019 & WiFeS/ANU 2.3m & 7000  & $29~000       $ & $4.66       $ & $-2.72       $ & [6] \\
ZGP-BLAP-09                & 19:14:40.84 & $+19$:38:25.6 & 17.16 & 23.264 & 2020 & OSIRIS/GTC     & 2500  & $     -      $ & $    -      $ & $     -      $ & [7] \\
ZGP-BLAP-01$^\ddag$        & 03:51:43.67 & $+58$:45:03.9 & 16.91 & 18.933 & 2021 & LRIS/Keck-I    & 1000  & $27~600\pm 400$ & $4.49\pm0.06$ & $-0.62\pm0.05$ & [8] \\
OGLE-BLAP-042              & 11:55:36.75 & $-64$:26:53.4 & 17.75 & 62.051 & 2023 & MagE/Magellan  & 4100  & $28~300\pm1000$ & $4.19\pm0.14$ & $-0.52\pm0.12$ & [5] \\
OGLE-BLAP-044              & 12:29:31.78 & $-60$:48:46.1 & 15.85 &  8.471 & 2023 & MagE/Magellan  & 4100  & $32~700\pm 200$ & $5.27\pm0.03$ & $-2.79\pm0.09$ & [5] \\
OGLE-BLAP-049              & 16:00:21.31 & $-57$:49:26.4 & 16.90 & 16.402 & 2023 & MagE/Magellan  & 4100  & $29~300\pm 400$ & $4.91\pm0.06$ & $-0.67\pm0.04$ & [5] \\
ZGP-BLAP-10                & 19:36:23.86 & $+05$:05:02.6 & 15.11 & 55.182 & 2023 & MagE/Magellan  & 4100  & $29~500\pm 700$ & $4.28\pm0.10$ & $-0.81\pm0.10$ & this work \\
\hline
\end{tabular}}
\end{center}
{\scriptsize
\begin{flushleft}
Other names: * =OGLE-BLAP-035, \dag ~=OGLE-BLAP-093, \ddag ~=TMTS-BLAP-1 \\
References: [1] Pietrukowicz \etal (2017), [2] Bradshaw \etal (2024), [3] Ramsay \etal (2022), [4] Kupfer \etal (2019), [5] Pietrukowicz \etal (2025), [6] Chang \etal (2023), \newline [7] McWhirter and Lam (2022), [8] Lin \etal (2023) \\
\end{flushleft}}
\end{sidewaystable}

McWhirter and Lam (2022) combined information from Gaia Data Release~2 (DR2; Gaia Collaboration 2018), ZTF DR3 (Bellm \etal 2019), and Pan-STARRS~1 (PS1; Magnier \etal 2020) to search for candidates for BLAPs at declinations $\delta > -15\arcd$. They identified 22 candidate BLAPs that were designated with a unique identification starting with ZGP, which comprises the first letter of the survey's name: ZTF, Gaia, and Pan-STARRS. Sixteen objects with periods in the range 17.0--55.2~min were classified as candidates for low-gravity BLAPs and named from ZGP-BLAP-01 to ZGP-BLAP-16. Six objects with shorter periods (2.4--8.2~min) were selected as candidates for high-gravity BLAPs and named from ZGP-HGBLAP-01 to ZGP-HGBLAP-06. According to McWhirter and Lam (2022), in their sample, one object is confirmed (ZGP-BLAP-09) and ten other pulsators are high-confidence BLAP candidates (including ZGP-BLAP-10, one of the investigated objects in this work). For all the variable stars, except ZGP-BLAP-03, the authors obtained (very) low-resolution spectra. However, in many cases, characteristic spectral features like helium lines could not be resolved and the stars require higher-resolution spectra in order to be confirmed as genuine BLAPs.

Among the reported 22 candidates for BLAPs in McWhirter and Lam (2022), three are located at latitudes $|b|>30\arcd$: ZGP-BLAP-03, ZGP-BLAP-04, and ZGP-BLAP-015. All the remaining candidates have $|b|<18\arcd$. So far, all but one BLAP have been discovered within $|b|<12\arcd$. The only exception is HD133729, which is located at $(l,b)=(333\zdot\arcd 96, +22\zdot\arcd 94)$ and a distance of less than 0.5 kpc, and is considered to be a BLAP orbiting a main-sequence B-type star with an orbital period of about 23.084~d (Pigulski \etal 2022).

The main goal of this work is to do a spectroscopic follow-up of the three high-latitude BLAP candidates to confirm, or not, the presence of BLAPs outside of the Galactic bulge and disk. The follow-up also includes the low-latitude but unconfirmed object ZGP-BLAP-10\footnote{This object was selected due to its availability from Las Campanas Observatory in bright time in June/July.}. In the following, we present our four target stars and their photometric properties in Sect.~2. In Sect.~3, we present the details of our spectroscopic follow-up and the updated classification of the BLAP candidates, and we draw a short conclusion in Sect.~4. 


\Section{Photometric Properties of the Target Candidates for BLAPs}

Basic observational parameters of the four investigated candidates for BLAPs from McWhirter and Lam (2022) are given in Table~2. In~Table~3, we provide additional information from Gaia DR3 (Gaia Collaboration 2021), including parallaxes. Fig.~1 presents phase-folded light curves in the $r$-band based on time-series data from ZTF DR22\footnote{https://irsa.ipac.caltech.edu/} covering years 2018--2024. The periods of the variables range from 46.7 to 55.2~min, while their full $r$-band amplitudes are between 0.13 and 0.22~mag. Objects ZGP-BLAP-03, ZGP-BLAP-04, and ZGP-BLAP-15 are located at Galactic latitudes $|b|>30\arcd$ and have mean $r$-band brightness between 18 and 19~mag. Their phased light curves have a nearly sinusoidal shape with a mild asymmetry and rounded minima, typical for short-period pulsating stars like those of $\delta$~Sct type (Pietrukowicz \etal 2020). Object ZGP-BLAP-10, with a period of 55.182 min, is the brightest of the four candidates with $r\approx15.2$~mag and is located at a lower galactic latitude ($b=-7\zdot\arcd61$). Its light curve has a sharp maximum and a flat minimum, and is very similar to the light curves of OGLE-BLAP-024 ($P=48.268$ min) and OGLE-BLAP-027 ($P=48.018$ min, Pietrukowicz \etal 2025). It is the only BLAP candidate with a Gaia parallax determination better than $3\sigma$. We note that we searched the time-series data for additional periods using the analysis of variance (Schwarzenberg-Czerny 1996) up to a frequency of 50 cycles per day, but nothing was detected above an amplitude of 0.002~mag. Hence, the four variable stars are very likely single-mode pulsators. In Fig.~2, we present color-magnitude diagrams with the positions of the variables. The diagrams were constructed using data from Gaia DR3 within $0.1\times0.1$ square-degree regions centered on the target stars. All four variables are located blueward of the main sequence observed in their directions.

\begin{table}[h!]
\centering
\caption{\small Coordinates, mean magnitudes, amplitudes, and periods of the verified objects}
\medskip
{\footnotesize
\begin{tabular}{lccccccc}
\hline
Designation &  RA(2000.0) &  Dec(2000.0)  &    $l$   &    $b$   & $r_{\rm ZTF}$ & amp$_r$ & $P$ \\
            &             &               &   [deg]  &   [deg]  & [mag]         & [mag]   & [min] \\
\hline
ZGP-BLAP-03 & 11:19:04.05 & $+13$:52:05.9 & $239.92$ & $+64.70$ & 18.96         & 0.22    & 53.704 \\
ZGP-BLAP-04 & 14:17:50.37 & $+00$:58:50.0 & $344.94$ & $+56.63$ & 18.33         & 0.17    & 46.681 \\
ZGP-BLAP-10 & 19:36:23.86 & $+05$:05:02.6 &  $42.65$ &  $-7.61$ & 15.25         & 0.17    & 55.182 \\
ZGP-BLAP-15 & 21:22:17.03 & $-00$:19:37.4 &  $52.02$ & $-33.16$ & 18.82         & 0.13    & 51.073 \\
\hline
\end{tabular}}
\end{table}

\begin{table}[h!]
\centering
\caption{\small Selected information on the verified objects from Gaia DR3}
\medskip
{\footnotesize
\begin{tabular}{lcccc}
\hline
Designation & Source ID           &   G   &   Parallax    & Variability class \\
            &                     & [mag] &     [mas]     & \\
\hline
ZGP-BLAP-03 & 3966636493833046400 & 18.96 & $0.01\pm0.25$ & not found to be variable \\
ZGP-BLAP-04 & 3666098876010103936 & 18.29 & $0.29\pm0.18$ & DSCT$|$GDOR$|$SXPHE \\
ZGP-BLAP-10 & 4291239234350458624 & 15.11 & $0.20\pm0.03$ & not found to be variable \\
ZGP-BLAP-15 & 2686382334320877952 & 18.78 & $0.17\pm0.20$ & not found to be variable \\
\hline
\end{tabular}}
\end{table}

\begin{figure}[h!]
\centerline{\includegraphics[angle=0,width=125mm]{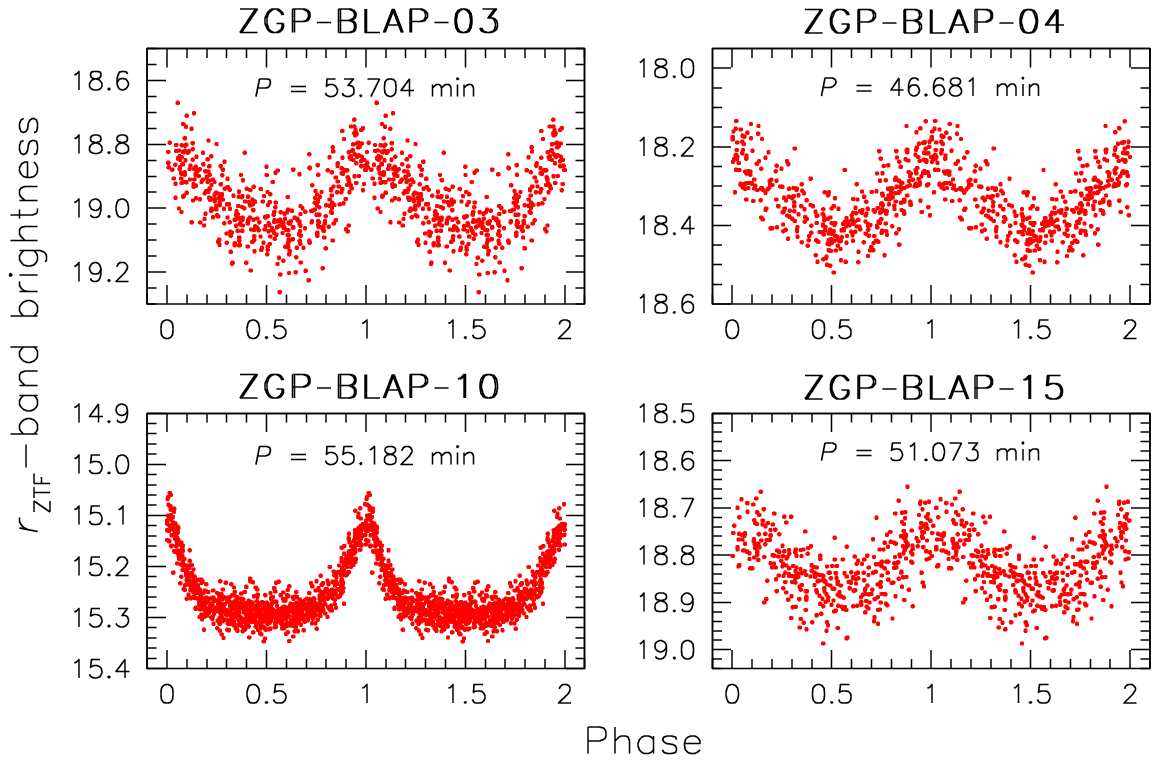}}
\FigCap{Phase-folded light curves in the ZTF $r$-band of the investigated variable stars.}
\end{figure}

\begin{figure}[h!]
\centerline{\includegraphics[angle=0,width=125mm]{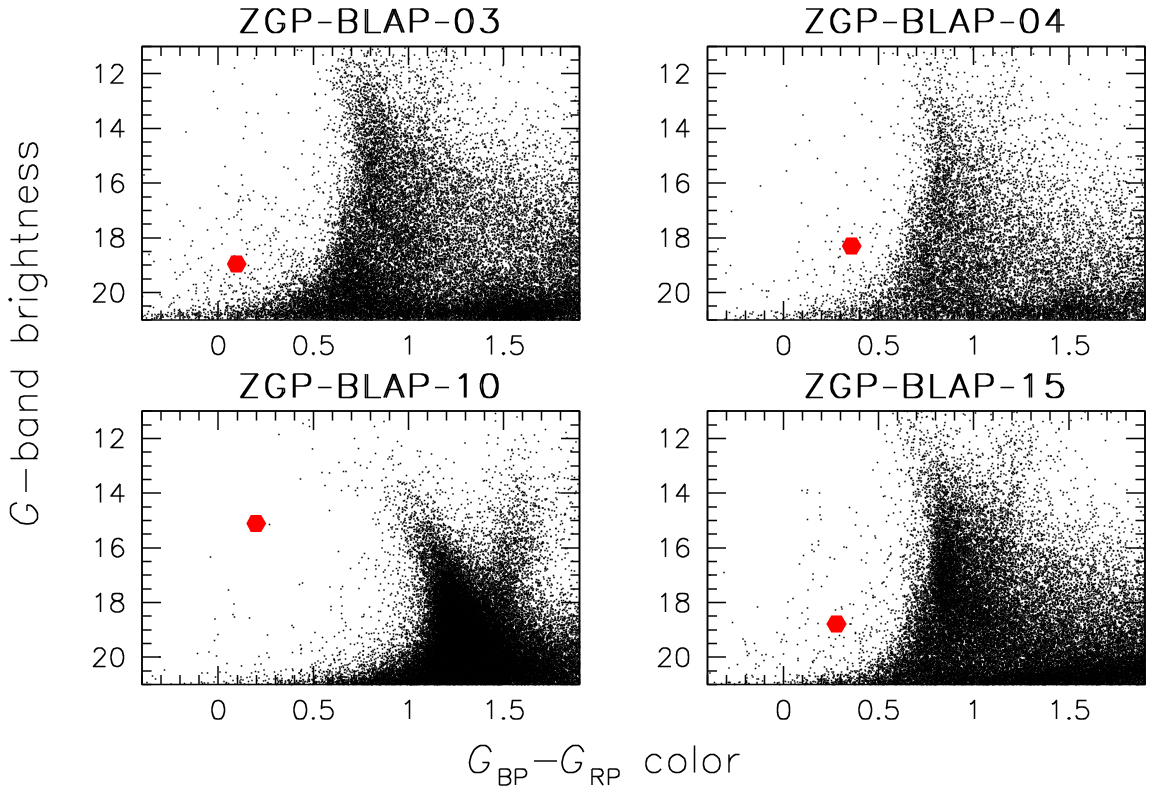}}
\FigCap{Color-magnitude diagrams with the positions of the investigated variable stars (red circles). The black dots are stars found in Gaia DR3 in a 0.1 $\times$ 0.1 square-degree regions centered on the BLAP candidates.}
\end{figure}


\Section{Spectroscopic Observations}

\begin{figure}[h!]
\centerline{\includegraphics[angle=90,width=125mm]{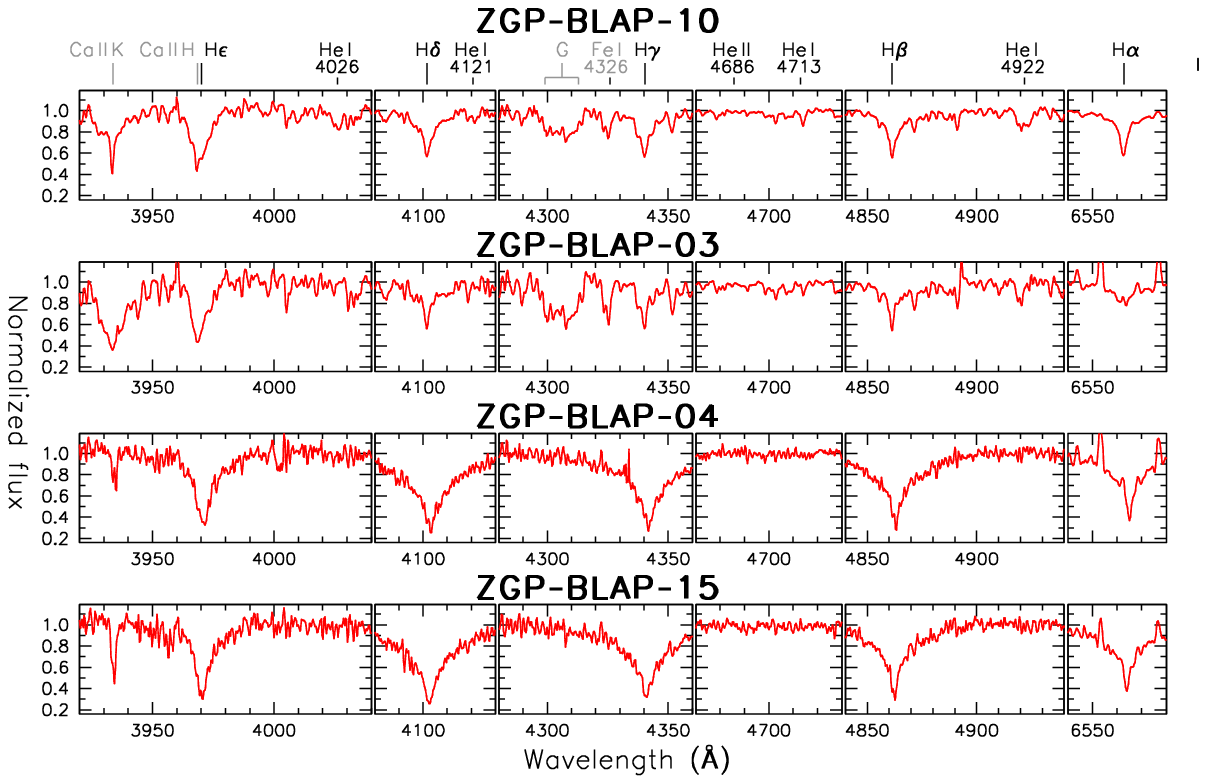}}
\FigCap{MagE/Magellan spectra of the investigated variable stars. Note that He~\textsc{i} lines are present only in the low-latitude object ZGP-BLAP-010.}
\end{figure}

We obtained moderate-resolution spectra of the four BLAP candidates using the Magellan Echellete (MagE) Spectrograph attached to the 6.5-m Magellan/Baade telescope located at Las Campanas Observatory (LCO), Chile. LCO is operated by the Carnegie Institution for Science. The observations were carried out in bright time during engineering nights in the middle of 2023. An observing log is presented in Table~4. MagE covers a wavelength range of 3200--10~000~\AA. The spectra were taken at a slit width of $1\zdot\arcs0$ and $1\times1$ binning readout, which provided us a resolving power of $R\approx4100$ or measured spectral resolution of 1.0~\AA ~at 4000~\AA. The slit was aligned to the parallactic angle. For each object, we took two exposures of the same length to cover one pulsation cycle. The scientific exposures were followed by an exposure of the Th-Ar lamp. The spectra were reduced using the IRAF package\footnote{IRAF was distributed by the National Optical Astronomy Observatory, USA, which is operated by the Association of Universities for Research in Astronomy, Inc., under a cooperative agreement with the National Science Foundation.}. Debiasing, flat-fielding, spectrum extraction, and wavelength calibration were performed in the standard way. We note that the spectra presented in this work were obtained during the same engineering run as the spectra of OGLE-BLAP-042, OGLE-BLAP-044, and OGLE-BLAP-049 presented in Pietrukowicz \etal (2025).

\begin{table}[h!]
\centering
\caption{\small Log of observations conducted with the MagE spectrograph}
\medskip
{\footnotesize
\begin{tabular}{llccl}
\hline
Designation & Observing night   & Exposure start  & $t_{\rm exp}$  & Sky quality \\
            &                   & BJD$_{\rm TDB}$ &      [s]       & \\
\hline
ZGP-BLAP-03 & 2023 Jun 3/4      &  2460099.49933  & $2\times 1610$ & cirrus clouds \\
ZGP-BLAP-04 & 2023 Jul 31/Aug 1 &  2460157.46247  & $2\times 1400$ & clear sky \\
ZGP-BLAP-10 & 2023 Jun 3/4      &  2460099.79845  & $2\times 1655$ & cirrus clouds \\
ZGP-BLAP-15 & 2023 Jul 29/30    &  2460155.67954  & $2\times 1532$ & clear sky \\
\hline
\end{tabular}}
\end{table}


\Section{Results of the Spectroscopic Follow-up}

In Fig.~3, we present the normalized spectra of the investigated variable objects over the wavelength ranges featuring the Balmer series and helium lines typical of BLAPs. A simple inspection of the spectra of ZGP-BLAP-04 and ZGP-BLAP-15 shows that the only visible features in these stars are strong hydrogen lines and the Ca~\textsc{ii}~K line. We found that the spectra could be reproduced by model atmospheres on both temperature sides of the Balmer lines maximum: either at $T_{\rm eff} \approx 13~000$~K with a low helium abundance, or at $T_{\rm eff} \approx 7500$~K with a low metallicity. The model atmospheres and fitting technique used for these two stars are presented in Irrgang \etal (2018). To discriminate between both solutions, we used the magnitudes of the stars, retrieved from various catalogs (see the caption of Fig.~5), to construct their spectral energy distribution (SED). The fits of the SED were made following the method described in Heber \etal (2018). Given the low reddening ($E(B-V)<0.05$ mag) expected in the direction of these two stars from reddening maps (Schlegel \etal 1998, Schlafly and Finkbeiner 2011), the SEDs of both stars are better reproduced with the cold solution. Thus, with $T_{\rm eff}$ of 7500~K and a metallicity [Fe/H] $\lesssim-2.5$ dex, ZGP-BLAP-04 and ZPG-BLAP-15 are evident SX~Phe-type pulsators, which are Population II counterparts of $\delta$~Sct stars. The small parallaxes at high galactic latitudes indicate that the two stars are halo objects.

The spectra of the remaining two objects, ZGP-BLAP-03 and ZGP-BLAP-10, show features characteristic for a solar spectrum, such as the G-band, Fe~\textsc{i} $\lambda4326$ and Fe~\textsc{i} $\lambda4920$ lines. This stems from the fact that the spectra were taken through cirrus clouds illuminated by the full Moon. In both objects, the Balmer lines are also present, but they are weaker than in the two stars discussed previously. The spectrum of ZGP-BLAP-03 does not show helium lines and the SED of this object can be reproduced by a single cool star with $T_{\rm eff}=7300\pm500$~K. Although we cannot estimate the metallicity of this object, its very high galactic latitude and a large distance (small parallax) indicate that it is another SX~Phe-type variable residing in the Galactic halo.

Finally, a closer inspection at the spectrum of ZGP-BLAP-10 reveals the presence of He~\textsc{i} lines among the plethora of metallic (solar) features. Fitting both the spectrum and the SED leads to atmospheric parameters for the hot component that are typical of low-gravity BLAPs: $T_{\rm eff}=29~500\pm700$~K, $\log g = 4.28\pm0.10$ dex, and $\log (N_{\rm He}/N_{\rm H})=-0.81\pm0.10$ dex. In Fig.~4 and Fig.~5, we present the fits to the spectrum and SED, respectively.

\begin{figure}[h!]
\centerline{\includegraphics[angle=90,width=80mm]{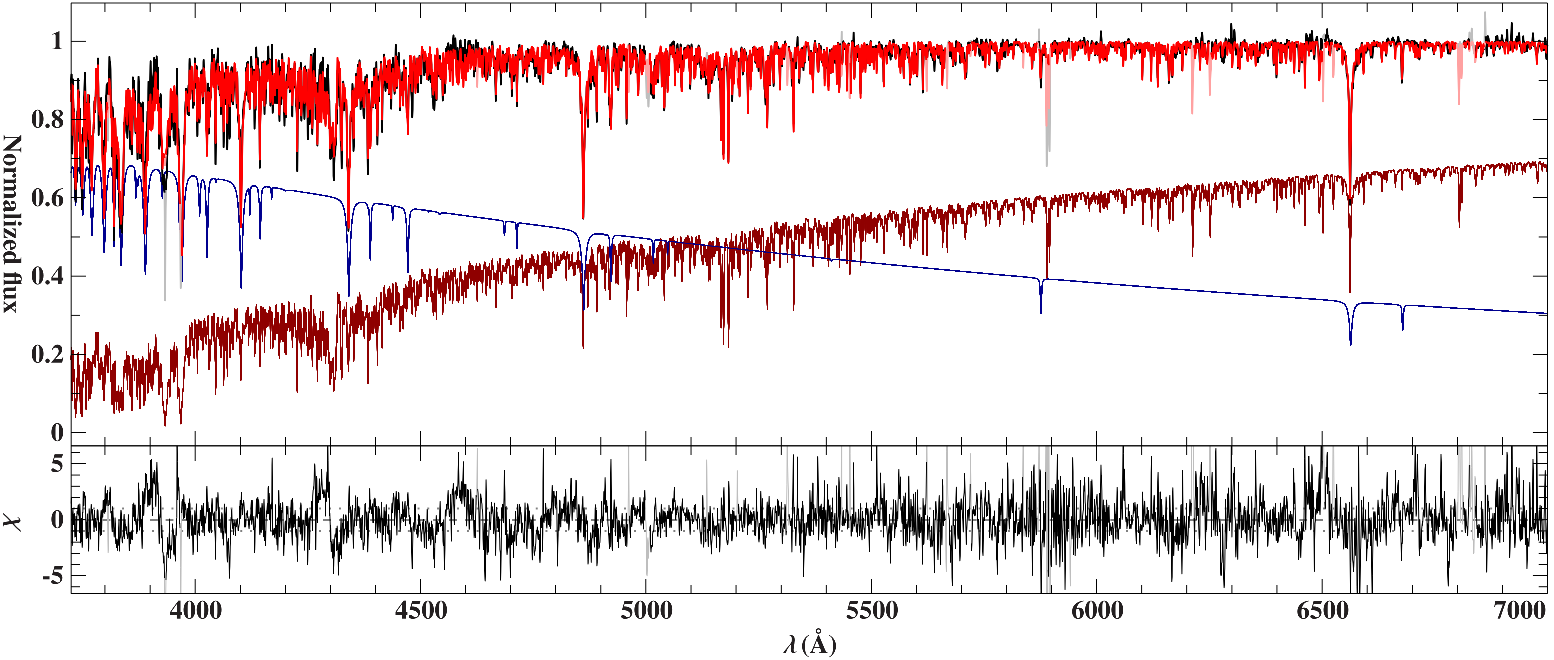}}
\FigCap{{\it Left panel:} Composite model fit to the normalized spectrum of ZGP-BLAP-10 (black line). The model for the hot component is shown in blue, the contribution of sunlight is plotted in dark red, and the combined model in strong red. {\it Right panel:} Uncertainty-weighted residuals.}
\end{figure}

\begin{figure}[h!]
\centerline{\includegraphics[angle=0,width=120mm]{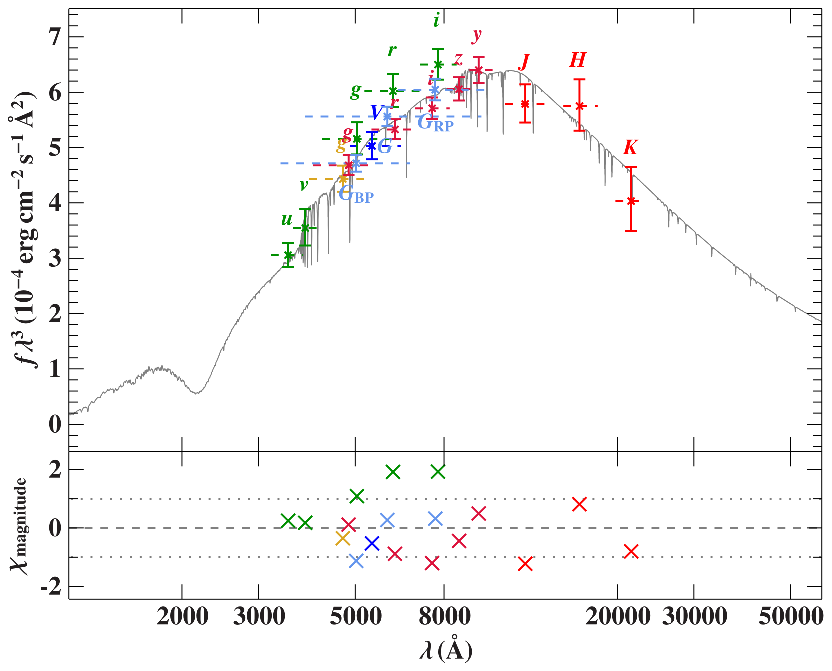}}
\FigCap{{\it Top panel:} SED fit for object ZGP-BLAP-10. The $uvgri$-band measurements (green points) were taken from the SkyMapper Southern Survey (SMSS) DR4 (Onken \etal 2024), $grizy$-band measurements (dark red points) from the Pan-STARRS1 survey DR2 (Chambers \etal 2016), $g$-band (yellow point) and $V$-band (navy blue point) measurements from the AAVSO Photometric All Sky Survey (APASS) DR9 (Henden \etal 2016), $G_{\rm BP}$, $G$, and $G_{\rm RP}$ measurements (pale blue points) from Gaia DR3 (Gaia Collaboration \etal 2021), $JHK$-band measurements (strong red points) from the Two Micron All-Sky Survey (2MASS, Skrutskie \etal 2006). {\it Bottom panel:} Uncertainty-weighted residuals.}
\end{figure}


\Section{Conclusions}

Our moderate-resolution spectra show that the variable objects ZGP-BLAP-03, ZGP-BLAP-04, and ZGP-BLAP-15 lack helium lines characteristic for hot stars. The objects are either of late spectral type A or early type F. Their periods of about 50~min and amplitudes of 0.1--0.2~mag are consistent with those observed in SX~Phe-type variables (\cf Holdsworth \etal 2014, Murphy \etal 2019, Soszy\'nski \etal 2021, Pietrukowicz \etal 2022). The presence of He~\textsc{i} lines in ZGP-BLAP-10 points to a hot star. Fitting both the spectrum and the SED confirms that this is a genuine BLAP. The light curve shapes of the four variable objects support this classification.

The follow-up observations presented in this work are summarized in Table~5 and may have some implications regarding the origin of BLAPs. All known pulsators of this type are found, so far, close to the Galactic plane. All genuine BLAPs except the brightest, relatively nearby object HD133729 are located within a stripe of $|b|<12\arcd$. The pulsators are not observed in globular clusters and have not been found in the Magellanic Clouds (Pietrukowicz 2018). BLAPs remain absent in metal-poor environments, indicating that metallicity might be a key factor in their formation mechanism.

\begin{table}[h!]
\centering
\caption{\small Spectroscopic features and true variability types of the investigated stars}
\medskip
{\footnotesize
\begin{tabular}{llccc}
\hline
Designation & Features in the stellar spectrum   & $T_{\rm eff}$ & True variability type \\
            &                                    & [K]           & \\
\hline
ZGP-BLAP-03 & Balmer series                      & $7300\pm500$  & SX~Phe \\
ZGP-BLAP-04 & Balmer series                      & $7540\pm25$   & SX~Phe \\
ZGP-BLAP-10 & Balmer series, He~\textsc{i} lines & $29~500\pm700$ & BLAP \\
ZGP-BLAP-15 & Balmer series                      & $7580\pm40$   & SX~Phe \\
\hline
\end{tabular}}
\end{table}


\Acknow{
P.P. has been supported by the Polish IDUB "Nowe Idee 3B" grant and "Microgrants" from the University of Warsaw, Poland. M.L. acknowledges funding from the Deutsche Forschungsgemeinschaft (grant LA 4383/4-1). In this work, we used time-series data from the Zwicky Transient Facility project. ZTF is supported by the National Science Foundation under grant no. AST-1440341 and a collaboration including Caltech, IPAC, the Weizmann Institute for Science, the Oskar Klein Center at Stockholm University, the University of Maryland, the University of Washington, Deutsches Elektronen-Synchrotron and Humboldt University, Los Alamos National Laboratories, the TANGO Consortium of Taiwan, the University of Wisconsin at Milwaukee, and Lawrence Berkeley National Laboratories. Operations are conducted by COO, IPAC, and UW. We also used data from the European Space Agency (ESA) mission Gaia, processed by the Gaia Data Processing and Analysis Consortium (DPAC). Funding for the DPAC has been provided by national institutions, in particular the institutions participating in the Gaia Multilateral Agreement.}


\end{document}